\documentclass[preprint,prb,showpacs,superscriptaddress]{revtex4}
\usepackage{amssymb,graphicx}

\begin{document}
\title{Theoretical determination of the necessary conditions for the formation of ZnO nanorings and nanohelixes}

\author{Z. C. Tu}
\affiliation{Computational Materials Science Center, National
Institute for Materials Science, Tsukuba 305-0047, Japan}
\author{Q. X. Li}
\affiliation{Computational Materials Science Center, National
Institute for Materials Science, Tsukuba 305-0047, Japan}
\affiliation{Hefei National Laboratory for Physical Sciences at
Microscale, University of Science and Technology of China, Hefei,
Anhui 230026, P. R. China}
\author{X. Hu}
\affiliation{Computational Materials Science Center, National
Institute for Materials Science, Tsukuba 305-0047,
Japan}
\begin{abstract}
The formation of ZnO nanorings and nanohelixes with large polar
surfaces observed in experiments [Nano Lett. \textbf{3}, 1625
(2003); J. Am. Chem. Soc. \textbf{126}, 6703 (2004)] is shown to
be a result of the competition between elastic energy, spontaneous
polarization-induced surface energy, volume energy, and
defect-induced energy. It is found that nanorings and nanohelixes
observed in experiments are stable and energetically favorable
structures.

\pacs{62.25.+g, 77.65.-j}
\end{abstract}

\maketitle

\section{Introduction\label{sec-intro}}
Piezoelectric material produces electrical charges when subjected
to pressure, and stretches or contracts when an electrical field
is applied, \cite{Nye} which has been widely used as acoustic
transducers in telephones, musical instruments, and so on.
Piezoelectric crystals have no center of symmetry, which induces
spontaneous polarization.\cite{Posternak} ZnO is a typical
piezoelectric material with Wurtzite structure.\cite{wangzljpc}
Zn$^{2+}$ and O$^{2-}$ planes arranging alternately along its
$c$-axis results in spontaneous polarization along this direction
that induces divergent surface energy of Zn-(0001) and
O-(000\={1}) surfaces. To maintain them stable, some researchers
propose that the surfaces are not reconstructed, in stead about
one-fourth charges transfer between them.
\cite{Taskerjpc,Noguerajpc,Wander01} However, theoretical
calculation and experiments reveal that the reconstruction is
preferable to charge transfer \cite{Kunat,Staemmler} in reducing
the surface energy.

After the discovery of ZnO nanobelts, \cite{Panzw} Kong and Wang
synthesized novel ZnO nanobelts, especially, nanorings and
nanohelixes whose surfaces were dominated by polarized Zn-(0001)
and O-(000\={1}) facets. \cite{Kongxy} They attributed the
observed phenomena as a result of surface polar charges and
explained that the formation of these nanostructures as a
consequence of minimizing the total energy contributed by
spontaneous polarization and elasticity. \cite{Kongxy} This model
requires the ratio of the thickness $t$ to the radius $R$ of the
nanorings to be smaller than a constant, and later more
experimental data located in the green domain of Fig.5a in
Ref.~\onlinecite{Hugheswl} supported this model. However, the
optimal relation between the thickness and the radius of nanorings
(black line of Fig.5a in Ref.~\onlinecite{Hugheswl}) in their
concise model cannot fit well the experimental data. To improve
this relation, an elaborate theory considering more factors than
their model but remaining its main spirit is needed.

Additionally, because of the symmetry of Wurtzite structure, we
know that there are 5 independent elastic constants for bulk ZnO
that have been measured as $c_{11}=190$ GPa, $c_{12}=110$ GPa,
$c_{13}=90$ GPa, $c_{33}=196$ GPa, $c_{44}=39$ GPa.
\cite{Azuhatat} From these experimental values, we see
$c_{11}\approx c_{33}$, $c_{12}\approx c_{13}$, and $c_{44}\approx
(c_{11}-c_{12})/2$. That is, bulk ZnO can be roughly regarded as
an isotropic material, and thus its Young's modulus can be
calculated as $\sim120$ GPa from these experimental data. It is
very surprising that Young's modulus of ZnO nanobelt is measured
experimentally only $\sim50$ GPa, \cite{Baixd,Maos} which is much
smaller than that of bulk ZnO. One needs to understand what makes
this large difference of Young's modulus between bulk ZnO and the
nanobelt.

In the present paper we try to address these issues. The rest of
this paper is organized as follows. In Sec.~\ref{sec-shape}, we
construct the shape formation energy of ZnO nanobelts that
consists of elastic energy, spontaneous polarization-induced
surface energy, volume energy, and defect-induced energy. And then
we put forward necessary conditions of structure existence. In
Sec.~\ref{sec-expl}, we use necessary conditions of structure
existence to explain nanorings and nanohelixes observed in
experiments. We find that they are stable and energetically
favorable structures. In Sec.~\ref{sec-sum}, we give a brief
summary and discussion. Some detailed derivations are supplied in
Appendix.

\section{Construction of shape formation energy\label{sec-shape}}
In order to quantitatively understand morphological problems of
nanostructures in theoretical point view, such as helixes of
multi-walled carbon nanotubes \cite{oycnt,zhangxb} and cholesterol
filaments, \cite{oybile,chungds} Ou-Yang and his co-authors
proposed a concept that the shape formation energy determines the
morphology of structures. Their concept can be summarized and
generalized as follows: a structure can exist if its shape
formation energy is minimum and less than zero, provided that the
thermal fluctuation can be neglected. We will define and construct
the shape formation energy of ZnO nanobelts in this section, and
then express explicitly the necessary conditions of structure
existence.

\subsection{Definition of shape formation energy}
The synthesis process of ZnO nanobelts is briefly described as
follows. \cite{Kongxy,Hugheswl} When ZnO powders are heated to
1350 $^{\circ}$C at a heating rate of 20 $^{\circ}$C/min, they
decompose into Zn$^{2+}$ and O$^{2-}$ at low pressure $10^{-3}$
torr. After a few minutes of evaporation and decomposition, the Ar
carrier gas is introduced at a flux of 25 standard cubic
centimeters per minute. The condensation products are carried to a
temperature zone of 400$\sim$500 $^{\circ}$C and deposit onto an
alumina substrate under the Ar pressure of 250 torr, where the end
products---nanobelts are formed. This process can be simplified as
a chemical ``reaction": $$\mathrm{Bulk ZnO}\rightleftharpoons
\mathrm{Condensation} \rightleftharpoons \mathrm{ZnO\
Nanobelts}.$$

\begin{figure}[!htp]
\includegraphics[width=6.5cm]{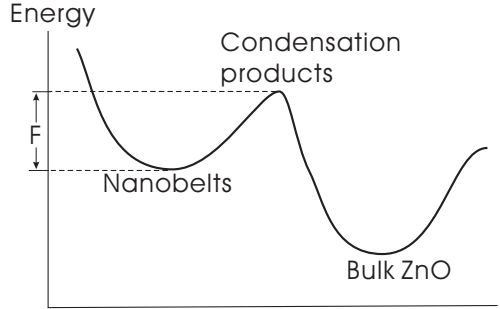}
\caption{\label{elandscape} Schematic energy landscape of the
synthesis process of ZnO nanobelts.}\end{figure}

Schematic energy landscape of the synthesis process of ZnO
nanobelts is shown in Fig.~\ref{elandscape}, where condensation
products are regarded as a meso-phase. The shape formation energy
$F$ is defined as the energy difference between ZnO nanobelt and
condensation products with the same number of atoms.

\begin{figure}[!htp]
\includegraphics[width=6.5cm]{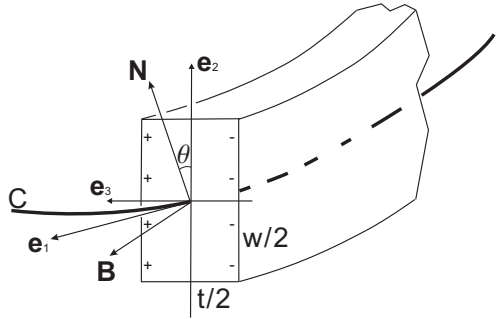}
\caption{\label{frame}Schematic figure of a ZnO nanobelt. $C$
represents its central line. $\mathbf{e}_1$ is the tangent vector
of curve $C$ that is perpendicular to the cross section of the
nanobelt. $\mathbf{e}_2$ and $\mathbf{e}_3$ (parallel to [0001]
direction of ZnO) are the principal axes of the cross section.
$\mathbf{N}$ and $\mathbf{B}$ are the normal and binormal vectors
of $C$, respectively. $\theta$ is the angle between $\mathbf{N}$
and $\mathbf{e}_2$.}\end{figure}

A ZnO nanobelt is a thin but long structure as schematically shown
in Fig.~\ref{frame}. The symbols $+$ and $-$ represent
Zn-terminated and O-terminated surfaces, respectively. The curve
$C$ is its central line; $\mathbf{e}_1$ is the tangent vector of
curve $C$ perpendicular to the cross section of the nanobelt;
$\mathbf{e}_2$ and $\mathbf{e}_3$ (parallel to [0001] direction of
ZnO) are principal axes of the cross section. $\mathbf{N}$ is the
normal vector of $C$ that determines the bending direction of $C$
in 3-dimensional space, and $\theta$ the angle between
$\mathbf{N}$ and $\mathbf{e}_2$. $\mathbf{B}$ is the binormal
vector of curve $C$. $w$ and $t$, the width and thickness of the
nanobelt because of $w\gg t$ in experiments,
\cite{Kongxy,Hugheswl} represent the dimensions in $\mathbf{e}_2$
and $\mathbf{e}_3$ directions, respectively. Because $w$ and $t$
are much smaller than the length of the nanobelt, the geometry of
the nanobelt is determined uniquely by its central line $C$ and
the angle $\theta$. Since the bending with $\theta=\pm \pi/2$ is
preferred to others because of $t\ll w$, we concentrate on this
case in this paper. The final form of the shape formation energy
of the nanobelt should depend merely on geometric quantities of
$C$, such as the curvature $\kappa$ and the torsion $\tau$. We
propose that it consists of elastic energy, surface energy, volume
energy, and defect-induced energy as discussed below.

\subsection{Elastic energy, surface energy, and volume energy}
The elastic energy per unit arc length can be written in an
invariant quadratic form including infinitesimal displacements of
the frame $\{\mathbf{e}_1,\mathbf{e}_2,\mathbf{e}_3\}$:
\begin{equation}E_c=\frac{k_{12}}{2}\left(  \frac{d\mathbf{e}_{1}}{ds}%
\cdot\mathbf{e}_{2}\right)  ^{2}+\frac{k_{13}}{2}\left[\left(  \frac{d\mathbf{e}%
_{1}}{ds}\cdot\mathbf{e}_{3}\right) ^{2}+\beta \left(
\frac{d\mathbf{e}_{2}}{ds}\cdot\mathbf{e}_{3}\right)
^{2}\right],\label{elaseng}\end{equation} where $ds$ is the arc
length element. $(d\mathbf{e}_{1}/ds)\cdot\mathbf{e}_{2}$ and
$(d\mathbf{e}_{1}/ds)\cdot\mathbf{e}_{3}$ represent, respectively,
the bending deformations around $\mathbf{e}_{3}$ and
$\mathbf{e}_{2}$, while $(d\mathbf{e}_{2}/ds)\cdot\mathbf{e}_{3}$
represents the torsion deformation around $\mathbf{e}_{1}$.
$k_{12}=Y_0tw^{3}/12$, $k_{13}=Y_0wt^{3}/12$, $\beta k_{13}$ with
$\beta=2/(1+\nu)$ are the bending and torsion rigidities,
\cite{Timoshenko} where $Y_0$ and $\nu$ are Young's modulus and
poisson ratio of ZnO. For normal materials $0\le\nu\le 1/2$,
$\beta$ is in the range between 4/3 and 2.

Consider the geometric relations $\mathbf{e}_{2}
=\mathbf{N}\cos\theta-\mathbf{B}\sin\theta$, $\mathbf{e}_{3}
=\mathbf{N}\sin\theta+\mathbf{B}\cos\theta$, and the Frenet
formula \cite{GrayA}
\begin{equation}
\left(\begin{array}{c}
d\mathbf{e}_{1}/ds\\
d\mathbf{N}/ds\\
d\mathbf{B}/ds
\end{array}\right) =\left(\begin{array}{ccc}%
0 & \kappa & 0\\
-\kappa & 0 & \tau\\
0 & -\tau & 0\end{array} \right)\left(\begin{array}{c}%
\mathbf{e}_{1}\\
\mathbf{N}\\
\mathbf{B}\end{array} \right),
\end{equation}
Eq.(\ref{elaseng}) is simplified as
\begin{equation}E_{c}=\frac{k_{13}}{2}\left[\left( \alpha\cos^{2}\theta+\sin
^{2}\theta\right) \kappa^{2}+\beta\left( \tau-d\theta/ds\right)
^{2}\right],\end{equation} where $\alpha=k_{12}/k_{13}$.
Especially, for $\theta=\pm \pi/2$, the elastic energy per unit
arc length can be written as:
\begin{equation}E_{c}=\frac{k_c}{2}\left( \kappa^{2}+\beta \tau
^{2}\right),\label{elastice}\end{equation} where
$k_c=k_{13}=Y_0wt^3/12$.

Because the nanobelt has two large polar surfaces, we must take
into account their surface energy. The surface energy per unit arc
length is phenomenologically expressed as
\begin{equation}E_s=2\gamma w,\end{equation} where $\gamma$ is a
constant quantity with the dimension of energy per unit area.

The volume energy comes from two sources: One is the Gibbs free
energy difference between the solid phase and the meso-phase
(condensation products); another is the electric energy induced by
the spontaneous polarization. The volume energy per unit arc
length is phenomenologically expressed as
\begin{equation}E_g=g_0wt,\label{volumee}\end{equation} where $g_0$ is a
constant quantity with the dimension of energy per unit volume.

\subsection{Defect-induced energy}
\begin{figure}[!htp]
\includegraphics[width=8cm]{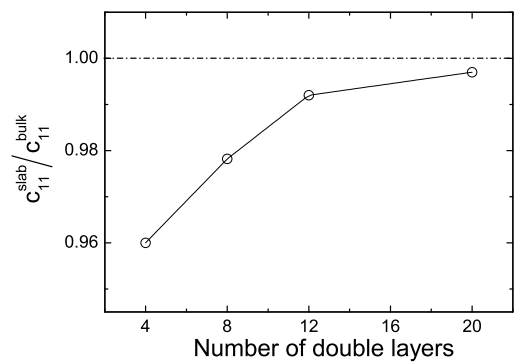}
\caption{\label{c11fig} Relative value of elastic constant
$c_{11}$ for a slab and bulk ZnO.}\end{figure}

As mentioned in Sec.~\ref{sec-intro}, the value of experimental
Young's modulus of nanobelts ($\sim$50 GPa) is less than half of
bulk ZnO's value ($\sim$ 120 GPa). In order to compare the elastic
constants of perfect ZnO thin films with bulk ZnO, we adopt a slab
model and density functional theory (DFT) method to calculate
them. \cite{DFTcal} To reduce the computational task, we take the
frozen-ion approximation and concentrate on the elastic constant
$c_{11}$. The computed ratio $c_{11}^{slab}$/$c_{11}^{bulk}$ as a
function of the slab thickness is shown in Fig.~\ref{c11fig}. It
is clear that the values of $c_{11}^{slab}$/$c_{11}^{bulk}$ are
close to 1, which implies that the mechanical properties of
perfect ZnO nanobelts have little difference from bulk ZnO. Thus
the nanobelts observed in experiments must contain defects that
reduce Young's modulus of nanobelts significantly.

There are several clues for the existence of defects in real
nanobelts. A direct one is high resolution images of nanobelts.
For example, the nanoring shown in Fig.4d of
Ref.~\onlinecite{Kongxy} is not an perfect, uniform structure.
High-resolution transmission electron microscopy also reveals that
planar defects exist in the nanobelts. \cite{wangsci,DingY}

It is hard to write an exact expression of defect-induced energy.
But if we only consider structures with $|\kappa t|\ll 1$ and
$|\tau t|\ll 1$, the defect-induced energy can be locally expanded
up to the second order terms of $\kappa t$ and $\tau t$. It is
expressed formally as
\begin{equation}E_d=-\left[A_1(\kappa t)^2+A_2(\tau t)^2+\mu (\kappa t)+A_3 \right]wt,\label{defecte1}\end{equation}
where $A_1$, $A_2$, $A_3$ and $\mu$ are constant quantities with
dimension energy per unit volume. To avoid the symmetry breaking
of chirality, the linear term of $\tau t$ is not included in the
above expression. Comparing Eq.(\ref{defecte1}) with
Eqs.(\ref{elastice}) and (\ref{volumee}), we find that $A_1$ and
$A_2$ play a similar role with the Young's modulus while $A_3$
influences the value of volume energy density $g_0$. The rest term
can induce a local spontaneous curvature of nanobelts randomly.
Recently, the idea that defects induce a local spontaneous
curvature has also been proposed to explain the cyclization
phenomenon of short DNA. \cite{yanjie}

\subsection{Necessary conditions of structure existence}
The final form of the shape formation energy is thus written as
\begin{eqnarray}
F&=&\int (E_c+E_s+E_g+E_d)ds\nonumber\\
&=&\int  \left[\frac{Yt^3}{24}(\kappa^{2}+\beta
\tau^{2})+(2\gamma+gt-\mu t^2\kappa)\right]w ds,\label{shapeng}
\end{eqnarray}
where $Y$ and $g$ are the effective Young's modulus and the volume
energy density of ZnO nanobelts, respectively. For example, the
value of $Y$ can be taken as the experimental value 50 GPa.

Now we estimate the effect of thermal fluctuation on the shapes of
nanobelts. Considering the typical growth temperature $T\sim 10^3$
K, the Young's modulus $Y=50$ GPa, the typical thickness $t\sim
10$ nm, and the width $w>50$ nm in the experiments,
\cite{Kongxy,Hugheswl} we estimate the persistence length of
nanobelts as $l_p=Ywt^3/12T\sim 1$ cm, \cite{Diom} which is much
larger than the typical length of nanobelts of several microns.
Thus the effect of thermal fluctuation can be neglected and the
necessary conditions of structure existence can be expressed as:
\begin{eqnarray}
&&\delta F=0,\label{detaf}\\
&&\delta^2 F\ge 0,\label{deta2f}\\
&& F\le 0,\label{fless0}
\end{eqnarray}
where $\delta F$ and $\delta^2 F$ represent the first and second
order variations of the shape formation energy, respectively. If a
structure satisfies the above three conditions, we call it stable
and energetically favorable structure.

From Eq.(\ref{detaf}), we arrive at the following shape equations
of nanobelts:
\begin{eqnarray}
&&
\kappa_{ss}+\frac{\kappa^{3}}{2}-\kappa\tau^{2}  +\frac{\beta}%
{2}\left[  4\tau\left(  \frac{\tau_{s}}{\kappa}\right)
_{s}+\frac{2\tau
_{s}^{2}}{\kappa}+3\kappa\tau^{2}\right]  +\frac{\chi \tau^2}{t}-\frac{(\eta+\xi t)\kappa}{2t^3}=0,\label{shap1}\\
&&  2\kappa_{s}\tau+\kappa\tau_{s}  +\beta\left[
\tau_{s}\tau^{2}-\left( \tau\kappa\right)  _{s}-\left(
\frac{\tau_s}{\kappa}\right) _{ss}\right]  -\frac{\chi
\tau_s}{t}=0,\label{shap2}\end{eqnarray} where $\chi=12\mu/Y$,
$\eta=48\gamma/Y$, and $\xi=24g/Y$. $()_s$ represents the
derivative respect to $s$. The detailed derivations of above two
equations are shown in Appendix. The necessary conditions of
structure existence are transformed into
Eqs.(\ref{deta2f})--(\ref{shap2}).

\section{Explanation to experiments\label{sec-expl}}
Now let us use Eqs.(\ref{deta2f})--(\ref{shap2}) to explain
nanorings and nanohelixes observed in experiments.
\subsection{Nanorings}
\begin{figure}[!htp]
\includegraphics[width=8cm]{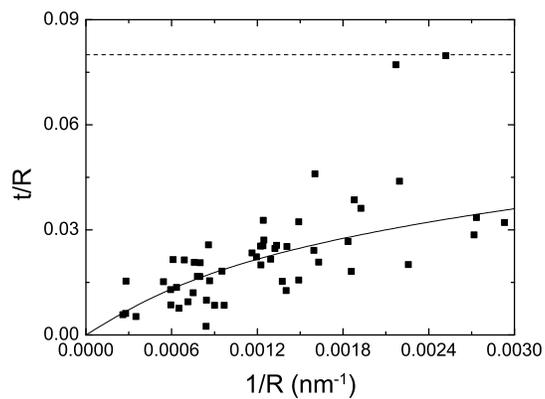}
\caption{\label{tdr-overR} Relation between $1/R$ and the ratio
$t/R$ of nanorings with radius smaller than 4000 nm. The data are
taken from Fig.5a of Ref.~\onlinecite{Hugheswl}. The solid line is
a fitting curve of the data by Eq.(\ref{nanoringe}). The dash line
is the supremum of $t/R$.}\end{figure}

For nanorings with radius $R$, one has $\kappa=1/R$ and $\tau=0$.
Eq.(\ref{shap1}) is transformed into
\begin{equation}(t/R)^{3} =\xi (t/R)+\eta/R,\label{nanoringe}\end{equation}
while Eq.(\ref{shap2}) is trivial. By fitting this equation to the
experimental data \cite{Hugheswl} as shown in
Fig.~\ref{tdr-overR}, we obtain $\xi=-0.001\pm 0.0007$ and
$\eta=0.03\pm0.01$ nm.

Now we test condition (\ref{deta2f}). In polar coordinate system
$(\rho,\phi)$, any perturbation in the vicinity of a cycle with
radius $R$ can be expressed as
$\rho=R(1+\sum_{n=-\infty}^{+\infty}a_ne^{in\phi})$, where
$a_{-n}=a^*_n$ $(n=0,1,2,\cdots,\infty)$ are small constants. The
second order variation is evaluated explicitly as
\begin{equation}
\delta^2 F=\sum_n (n^2-1)^2|a_n|^2Ywt^3/12R\ge 0
\end{equation}
under the validity of Eq.(\ref{nanoringe}).

Next, nanorings observed in experiments must satisfy condition
(\ref{fless0}). This requires $F=\pi Ywt^2(t/R-\chi)/6\le 0$ to be
valid for all experimental data. Thus we can take the supremum of
$t/R$ in Fig.~\ref{tdr-overR} as the value of $\chi$, i.e.,
$\chi=0.08$.

\subsection{Nanohelixes}
For a nanohelix with radius $r_0$ and pitch $p$, the pitch angle
is $\varphi=\arctan(p/2\pi r_0)$. The curvature and the torsion
are $\kappa=\cos^{2}\varphi/r_{0}$ and
$\tau=-\sin\varphi\cos\varphi/r_{0}$, respectively.
Eq.(\ref{shap1}) is transformed into
\begin{equation}
\cos^4\varphi+(3\beta-2)\sin^2\varphi\cos^2\varphi+\frac{2\chi
r_0}{t}\sin^2\varphi=\frac{r_0^2(\eta +\xi
t)}{t^3},\label{nanohelixe}
\end{equation}
while Eq.(\ref{shap2}) is trivial again. Eq.(\ref{nanohelixe})
sets a relation between the thickness, the pitch angle and the
radius of nanohelixes, from which we can estimate the thickness.
The results by using the data in Ref.~\onlinecite{Kongxy} are
summarized in Table \ref{helixtab}. These values are close to the
typical value 10 nm observed in experiments. \cite{Kongxy}

Additionally, the shape formation energy can be calculated as
\begin{equation}F=\frac{Ywt^3L}{12r_{0}^{2}}\left[\cos^{4}\varphi+\frac{2\beta-1}{4}\sin^{2}2\varphi-\frac{\chi r_{0}}{t}\cos2\varphi\right],\end{equation}
where $L$ is the total length of the nanohelixes. Its value is
listed in Table \ref{helixtab} for three nanohelixes observed in
the experiments using $\chi=0.08$. Moreover, we also found the
matrix
$$\left(\begin{array}{cc}\partial^2 F/\partial r_0^2 &
\partial^2 F/\partial r_0\partial \varphi\\
\partial^2 F/\partial r_0\partial \varphi &\partial^2 F/\partial\varphi^2
\end{array}\right)$$ is positive definite for these
nanohelixes. Therefore, these nanohelixes are stable and
energetically favorable structures.

\section{Conclusion\label{sec-sum}}
In the present work, we construct the shape formation energy of
ZnO nanobelts that consists of elastic energy, spontaneous
polarization-induced surface energy, volume energy, and
defect-induced energy. We put forward the necessary conditions of
structure existence and find that nanorings and nanohelixes of ZnO
observed in experiments are stable and energetically favorable
structures. We obtain an important result that nanorings and
nanohelixes must satisfy Eqs.(\ref{nanoringe}) and
(\ref{nanohelixe}), respectively. We notice that it is easy to
measure experimentally the values of the radius of nanorings or
the pitch and radius of nanohelixes, while it is difficult to
measure their thickness of about ten nanometers in high precision
because the error of InLens detector is $\pm 2$ nm.
\cite{Kongxy,Hugheswl} Eqs.(\ref{nanoringe}) and
(\ref{nanohelixe}) provide a way to overcome this difficulty. The
present theory may also be developed to explain the recent
synthesized superlattice-structured nanohelxes. \cite{gaopxsci05}

Because of the influence of defects, Young's modulus $Y$ and
volume energy density $g$ are quite different from the value of
perfect and uniform nanobelts. For example, $Y=50$ GPa is just
half of the value of perfect nanobelts. $g$ is estimated as
$\sim-10^6$ J/m$^3$ by $\xi=-0.001$. This value departs largely
from the cohesive energy $\sim-10^{10}$ J/m$^3$ of ZnO.
\cite{Cheny} The surface energy density $\gamma$ is estimated as
$0.03$ J/m$^2$, which cannot compare with the cleavage energy
$\sim4$ J/m$^2$ for perfect Zn-(0001) and O-(000\={1}) surfaces
obtained by \textit{ab initio} calculations. \cite{Wander01} There
is no theory on the parameter $\mu$ at present, which is expected
to depend on the defect density and experimental conditions. We
estimate its value $\mu\approx10^8$ J/m$^3$ from the experiment
published in Ref.~\onlinecite{Hugheswl}.

In the above discussion, we do not include flexoelectric effect
\cite{Meyer,Tagantsev} of piezoelectric materials in nanoscale.
This effect is too small for crystals \cite{Tagantsev} to
influence the above results in statics. But it may play a role in
the kinetic growth of ZnO nanobelt so that nanorings with
Zn-terminated and O-terminated inner surfaces have different
weights in the end products. Through analyzing the weights, one
may reveal the flexoelectric effect of ZnO in experiments.

\section*{Acknowledgement}
We are very grateful to Prof. Z. L. Wang, Dr. W. L. Hughes and X.
Y. Kong for kindly sending us experimental data. Calculations have
been performed on Numerical Materials Simulator (HITACHI SR11000)
at the Computational Materials Science Center, National Institute
for Materials Science.

\appendix
\section*{Appendix}
We present the derivation of Eqs.(\ref{shap1}) and (\ref{shap2})
from the first order variation of functional (\ref{shapeng}) under
the periodic boundary or fixed boundary. The basic idea is similar
to Ref.~\onlinecite{tzcjpa}. When we perform the variational
calculation, the volume and density of a ZnO nanobelt are allowed
to vary, while the total mass is conserved. However, the variation
is performed for infinitely small perturbation which has very
small effect on the elastic coefficients in Eq.(\ref{shapeng}).
The final Euler-Lagrange equations obtained from the variational
process including this effect are the same as those derived from
the variation with elastic coefficients fixed to constant.

Let $E_{1}=\mathbf{e}_{1}$, $E_{2}=\mathbf{N}$, $E_{3}=\mathbf{B}$
and $dE_{i} =w_{ij}E_{j}$, where Einstein's summation convention
is used and ``$d$" is an exterior differential operator. Thus we
have $d\mathbf{r}=E_{1}ds$, $w_{12}=\kappa ds,w_{13}=0,w_{23}=\tau
ds$ by considering the Frenet formula.

First, we consider $\delta\mathbf{r}=\Omega_{2}E_{2}$ and $\delta
E_i=\Omega_{ij}E_j$. Using the formula $d\delta\mathbf{r}=\delta
d\mathbf{r}$ and $d\delta E_{i}=\delta dE_{i}$, \cite{tzcjpa} we
have
\begin{eqnarray}
\delta ds  &=&\Omega_{2}w_{21}=-\kappa ds\Omega_{2},\\
\Omega_{12}ds  &  =&d\Omega_{2},\\
\Omega _{13}&=&\tau\Omega_{2},\\
\delta w_{ij}&=&d\Omega_{ij}+\Omega _{ik}w_{kj}-w_{ik}\Omega_{kj}.
\end{eqnarray}
Because $\delta w_{13}=0$, we obtain
\begin{equation}
\Omega_{23}ds =[2d\left(\tau\Omega_{2}\right)
-\Omega_{2}d\tau]/{\kappa}.
\end{equation}
Using above five equations, we can prove
\begin{eqnarray}
\delta\kappa ds  &=&d\Omega_{12}+\left( \kappa^{2}-\tau^{2}\right)
\Omega_{2}ds,\\
\delta\tau ds  &=&d\Omega_{23}+2\kappa\tau ds\Omega_{2}.
\end{eqnarray}
Considering Stokes theorem and above seven equations, we arrive at
\begin{eqnarray}
\delta\int\kappa^{2}ds&=&\int2\left(
\kappa_{ss}+\kappa^{3}/2-\kappa \tau^{2}\right)  \Omega_{2}ds,\\
\delta\int\tau^{2}ds&=&\int\left[  4\tau\left(
\frac{\tau_{s}}{\kappa}\right)  _{s}+\frac
{2\tau_{s}^{2}}{\kappa}+3\kappa\tau^{2}\right]  \Omega_{2}ds,\\
\delta\int \kappa ds&=&-\int \tau^{2}\Omega_{2}ds.
\end{eqnarray}
Consequently, we obtain
\begin{eqnarray}
\delta F&=&\frac{Ywt^3}{12}\int\left\{\left(
\kappa_{ss}+\frac{\kappa^{3}}{2}-\kappa\tau^{2}\right)
+\frac{\beta}{2}\left[ 4\tau\left( \frac{\tau_{s}}{\kappa}\right)
_{s}+\frac{2\tau_{s}^{2}}{\kappa}+3\kappa\tau^{2}\right]  \right\}
\Omega_{2}ds\nonumber\\
&+&\int w[\mu t^2\tau^{2}-(2\gamma+gt)]\kappa\Omega_{2}ds.
\end{eqnarray}
Because $\Omega_{2}$ is an arbitrary function, $\delta F=0$ gives
Eq.(\ref{shap1}).

Secondly, we consider $\delta\mathbf{r}=\Omega_{3}E_{3}$. Using
the formula $d\delta\mathbf{r}=\delta d\mathbf{r}$ and $d\delta
E_{i}=\delta dE_{i}$, we have
\begin{eqnarray}
\delta ds  &=&\Omega_{3}w_{31}=0,\\
\Omega_{12}&=&-\tau\Omega_{3},\\
\Omega_{13}ds &=&d\Omega_{3},\\
\delta w_{ij}&=&d\Omega_{ij}+\Omega _{ik}w_{kj}-w_{ik}\Omega_{kj}.
\end{eqnarray}
Because $\delta w_{13}=0$, we obtain
\begin{equation}\Omega_{23}ds =[d\Omega_{13}-\tau^{2}\Omega_{3}ds]/\kappa.\end{equation}
Using above five equations, we can prove
\begin{eqnarray}
\delta\kappa ds  &=&-2d\left( \tau \Omega_{3}\right)
+\Omega_{3}d\tau,\\
\delta\tau ds  &=&d\Omega_{23}+\kappa d\Omega_{3}.
\end{eqnarray}
Considering Stokes theorem and above seven equations, we arrive at
\begin{eqnarray}
\delta\int\kappa^{2}ds&=&\int\left(  4\kappa_s\tau+2\kappa\tau_s\right)  \Omega_{3}ds,\\
\delta\int\tau^{2}ds&=&=2\int\left[  \tau^{2}\tau_s-\left(
\tau\kappa\right) _{s}-\left(  \frac{\tau_s}{\kappa}\right)
_{ss}\right]  \Omega_{3}ds,\\
\delta\int \kappa ds&=&\int\tau_{s}\Omega_{3}ds.
\end{eqnarray}
Consequently, we obtain
\begin{eqnarray}
\delta F&=&\frac{Ywt^3}{12}\int
\left\{2\kappa_{s}\tau+\kappa\tau_{s} +\beta\left[
\tau_{s}\tau^{2}-\left( \tau\kappa\right) _{s}-\left(
\frac{\tau_s}{\kappa}\right) _{ss}\right]  -\frac{24\mu
\tau_s}{Yt}\right\}\Omega_{3}ds.
\end{eqnarray}
Because $\Omega_{3}$ is an arbitrary function, $\delta F=0$ gives
Eq.(\ref{shap2}).

\newpage

\begin{table}[!htp]
\caption{Radii, pitches, estimated thicknesses and shape energies
for different helixes. The data for radii and pitches are taken
form Ref.~\onlinecite{Kongxy}. $L$ represents the length of
helixes.\label{helixtab}}
\begin{ruledtabular}
\begin{tabular}{cccc}
$r_0$ (nm) & $p$ (nm) & $t$ (nm) & $F$ ($Ywt^3L/12r_0^2$)\\
\hline 342 & 222 & 12.4$\sim$12.5 & -1.1\\
175 & 133 & 8.5$\sim$8.6 & -0.6\\
240 & 380 & 9.6$\sim$9.8 & -0.7
\end{tabular}
\end{ruledtabular}
\end{table}

\end{document}